\documentclass[useAMS,usenatbib,usegraphicx]{mn2e}
\usepackage{amssymb,amsmath}
\usepackage{color}

\title[Stellar haloes of disc galaxies at z$\sim$1]{Stellar haloes of disc galaxies at z$\sim$1}
\author[Ignacio Trujillo and Judit Bakos]{Ignacio Trujillo$^{1,2}$\thanks{E-mail:
trujillo@iac.es} and Judit Bakos$^{1,2}$\footnotemark[1]\\
$^{1}$ Instituto de
Astrof\'isica de Canarias, V\'ia L\'actea s/n, 38200 La Laguna, Tenerife, Spain\\
$^{2}$ Departamento de Astrof\'isica, Universidad de La Laguna, E-38205 La Laguna, Tenerife, Spain}
\begin{document}

\date{Accepted 1988 December 15. Received 1988 December 14; in original form 1988 October 11}

\pagerange{\pageref{firstpage}--\pageref{lastpage}} \pubyear{2002}

\maketitle

\label{firstpage}

\begin{abstract}

Taking advantage of the ultradeep  near-infrared imaging obtained with the Hubble Space Telescope on the Hubble Ultra
Deep Field, we detect and explore for the first time the properties of the stellar haloes of two
Milky Way-like galaxies at z$\sim$1.  We find that the structural properties of those haloes (size and
shape) are similar to the ones found in the local universe. However, these high-z stellar haloes are
approximately three magnitudes brighter and exhibit bluer colours ((g-r)$\lesssim$0.3 mag) than their local
counterparts. The stellar populations of z$\sim$1 stellar haloes are compatible with having ages
$\lesssim$1 Gyr. This implies that the stars in those haloes were formed basically at 1$<$z$<$2. This
result matches very well the theoretical predictions that locate most of the formation of the stellar
haloes at those early epochs. A pure passive evolutionary scenario, where the stellar populations of
our high-z haloes simply fade to match the stellar halo properties found in the local universe, is
consistent with our data.

\end{abstract}


\section{Introduction}

A natural prediction of the hierarchical galaxy formation scenario  is the ubiquitous presence of
stellar haloes surrounding the galaxies (e.g. Eggen et al. 1962; Searle \& Zinn 1978; Steinmetz \&
M\"uller 1995; Bekki \& Chiba 2001; Samland \& Gerhard 2003). These extended and diffuse stellar
components are formed from the debris of disrupted satellites accreted along the cosmic time (e.g.
Brook et al. 2003; Bullock \& Johnston 2005; Abadi et al. 2006). Recent cosmological simulations
predict that the amount of stellar mass in these haloes should be
$\sim$10$^{8}$-10$^{9}$M$_{\sun}$ for Milky Way (MW)-like objects. If these simulations are
correct, most of the stellar halo mass would be assembled before z$\sim$1 and we would expect since
then a simple passive evolution of the stellar populations of these haloes towards the present (e.g.
Cooper et al. 2010; Font et al. 2011).

From the observational point of view, the detection of  stellar haloes is a major challenge.   For
example, Morrison (1993) estimates the surface brightness of the Galactic halo at the solar radius to
be V$\sim$27.7 mag/arcsec$^2$. In addition to the well-characterized MW halo (see a recent
review by Helmi 2008), deep surveys of M31 (e.g. Ferguson et al. 2002; Irwin et al. 2005; Kalirai et
al. 2006; Ibata et al. 2007; McConnachie et al. 2009) have revealed an extensive halo (to $\sim$150
kpc) with abundant substructure. M33 seems also to have a halo similar to the MW and M31,
despite its smaller total mass (McConnachie et al. 2006). Beyond the Local Group, there is growing
observational evidence showing that stellar haloes are ubiquitous and diverse. These studies have been
conducted in nearby galaxies by using resolved star count techniques (e.g. Mouhcine et al. 2005, 2007,
2010; Ibata et al. 2009; Radburn-Smith et al. 2011) or by extremely deep integrated photometry
observations able to detect low surface brightness features (e.g. Sackett et al. 1994; Shang et al.
1998;  Mart\'inez-Delgado et al. 2008, 2009; Jablonka et al. 2010; Bakos \& Trujillo 2012). Zibetti et
al. (2004; see also Bergvall et al. 2010), stacking 1047 galaxies, were able to observe the properties
of stellar haloes up to z$\sim$0.05 ($\sim$200 Mpc in distance). Finally, Zibetti \& Ferguson (2004),
using the Hubble Ultra Deep Field (HUDF; Beckwith et al. 2006), detected the stellar halo of a disc
galaxy at z=0.32. This last observation represents the current farthest detection of these faint
components of the galaxies.

The observed number and physical properties of the stellar streams in  nearby galaxies are in general agreement with the predictions from
the $\Lambda$ Cold Dark Matter  model (e.g. Bell et al. 2008; Gilbert et al. 2009, McConnachie et al. 2009; Starkenburg et al. 2009;
Mart\'inez-Delgado et al. 2010). Even the large discrepancies in the properties of the stellar haloes of similar mass disc galaxies as
the MW and M31 have been explained, within the cosmological context, as inherent to the system-to-system scatter in the halo
formation histories of these objects (e.g. Font et al. 2011). To gain, consequently, a deeper understanding of the formation mechanisms
of the disc galaxies is necessary: (a) to explore systematically the stellar haloes of nearby galaxies to even fainter surface
brightness magnitudes (V$>$30 mag/arcsec$^2$) where a plethora of substructures is expected to appear and (b) to probe the stellar
haloes of high redshift disc galaxies, where the properties of the stellar haloes are caught in earlier episodes of assembly. This paper deals with the second possibility. In particular, if the stellar halo formation scenarios are correct, a cosmic epoch
that it is worth exploring is z$\sim$1 where, as  mentioned before, the haloes should be already in place.

If observing stellar haloes in the local universe is difficult, the observation of stellar haloes at
z$\sim$1 requires the deepest ever observations from the space. At z=1, the cosmological dimming is
$\sim$3 mag/arcsec$^2$ and the optical red side of stellar emission (which contains most of the
stellar flux) is only visible using near-infrared (NIR) data. Fortunately, the observations to start
exploring the stellar haloes up to z$\sim$1 have been recently taken by using the Wide Field Camera 3
(WFC3) onboard the Hubble Space Telescope (HST) imaging the HUDF. In this pilot
paper, we present for the first time the stellar haloes of two MW-like galaxies at z$\sim$1. We will
show that these stellar haloes, in agreement with theoretical expectations, were much brighter and
bluer than present-day stellar haloes of similar mass disc galaxies.

The structure of this paper is as follows. In Section 2 we describe the data used in this paper, the
selection of our sample and the profiles extraction, and in Section 3 we present a detailed analysis
of our results. Finally, in Section 4, we perform a short discussion and a summary of our work. The
paper is complemented with an extensive Appendix  where we expand on technical details used at
conducting this article.

All magnitudes in this paper are given in the AB system unless otherwise stated. Throughout we assume a flat -dominated cosmology
($\Omega_m$=0.3, $\Omega_\Lambda$=0.7, and H$_0$=70 km s$^{-1}$ Mpc$^{-1}$).

\section[]{Data and sample selection}

Due to the faintness of the features we aim to explore in this paper, the study presented here
requires the nowadays state-of-the-art imaging in terms of depth and resolution. This implies using
the deepest optical and near-infrared imaging ever taken with the HST. The HUDF
(R.A.=03:32:39.0,  Dec = -27:47:29.1 (J2000)) is the point on the sky with the best HST data for this
project. 

\subsection{Data}

In the optical regime, the HUDF has been observed (Beckwith et al. 2006) with the WFC of the Advanced Camera for Surveys (ACS) in four filters: F435W (B), F606W (V), F775W (i) and
F850LP (z). The total exposure time in each band is as follows: 134880s (B), 135320s (V), 347110s (i)
and 346620s (z), and the AB mag. zeropoints reached are: 25.673 (B), 26.486 (V), 25.654 (i) and 24.862
(z).  Individual exposures in each filter were combined and drizzled to produce a single image per
band with the following characteristics: 10500$\times$10500 pixels with a pixel scale of 0.03
arcsec/pixel. These data were made public by the HUDF team at the following webpage:
http://archive.stsci.edu/pub/hlsp/udf/acs-wfc/.

In the NIR, the HUDF was observed both by the Near Infrared Camera and Multi-Object
Spectrometer (NICMOS) camera and more recently by the WFC3. The WFC3 observations are deeper and with
higher resolution than NICMOS; however, the WFC3 observations do not cover entirely the HUDF field.
For these reasons, in this paper, we have used both NIR data sets. The NICMOS HUDF (Thompson
et al. 2005) data consist on F110W and F160W drizzled images of 3500$\times$3500 pixels with a pixel
scale of 0.09 arcsec/pixel. The zeropoints are: 23.410 (F110W) and 23.220 (F160W). The images are
available at http://archive.stsci.edu/pub/hlsp/udf/nicmos-treasury/. The WFC3 HUDF imaging used here
(Bouwens et al. 2011) only covers the very central (4.7 arcmin$^2$) part of the HUDF but to an
extraordinary depth. The zeropoints are: 26.27 (F105W), 26.25 (F125W) and 25.96 (F160W). The combined
individual exposures in these filters were drizzled and have approximately 3000$\times$3000 pixels
with a pixel scale of 0.06 arcsec/pixel. The data is publicly available at:
http://archive.stsci.edu/prepds/hudf09/.

\subsection{Sample Selection}

The selection of our galaxies was conducted using the Rainbow Cosmological
Database\footnote{https://rainbowx.fis.ucm.es/} published by P\'erez-Gonz\'alez et al. (2008); see also
Barro et al. (2011a,b). This database is a vast compilation of photometric and spectroscopic data for
several of the deepest cosmological fields, such as GOODS-North and South, COSMOS, or the Extended
Groth Strip, among others. Using all the available photometry, they have built spectral energy
distributions (SEDs) covering the electromagnetic spectrum from the X-ray to the radio wavelengths.
Analysing these SEDs, they have derived very robust photometric redshifts and accurate estimates of
various stellar parameters (such as the stellar mass, the ultraviolet and infrared-based star formation rates, the
stellar population age, etc.).

We have selected from the Rainbow data base those objects with spectroscopic redshift within
0.8$<$z$<$1.2 and M$_{\star}$$>$5$\times$10$^{9}$M$_{\sun}$ (Kroupa 2001 IMF). These constraints
provide 15 galaxies. We have used the ACS z-band imaging to visually explore these objects. Two of
these galaxies are ellipticals and are not considered in our study on what follows. From the remaining
13 galaxies, 9 objects are irregular or appear distorted by interactions. There are four bona fide
spirals: UDF3372, UDF4438, UDF5417 and UDF9444. From those, we have finally selected only the two disc
galaxies with the lowest inclinations. This  mitigates potential effects of the dust on the surface
brightness profiles we have obtained. In addition, while exploring the outer stellar haloes, less
inclined galaxies are expected to be less affected  by point spread function (PSF) effects (see de Jong 2008) than edge-on
orientations.  The characteristics of our galaxies are shown in Table 1. We have also shown these
objects (in z and H bands) in Fig. 1. Unfortunately, UDF3372 is beyond the area covered by the HUDF
WFC3 pointing, so we have had to rely our analysis on the NIR using the NICMOS data.

\begin{figure*}
\includegraphics[width=16cm]{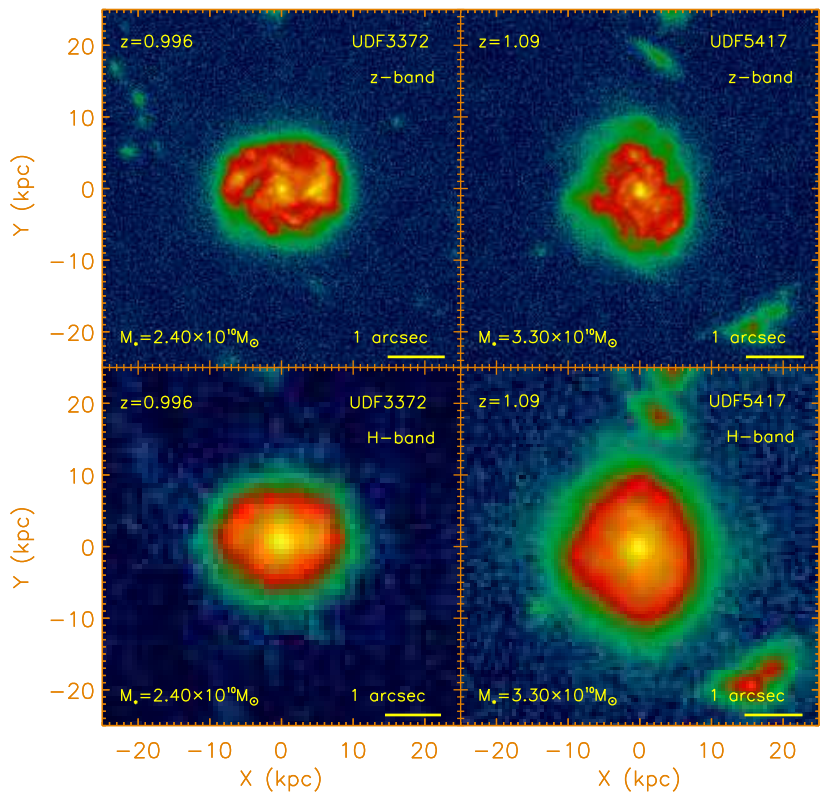}

\caption{Our sample of MW-like galaxies at z$\sim$1: UDF3372 and UDF5417. {\it Upper panels:} the ACS z-band ($\sim$B-band
rest frame) of the galaxies in our sample. {\it Bottom panels:} the NICMOS F160W band (UDF3372) and the WFC3 F160W band (UDF5417) ($\sim$I-band
restframe) of our galaxies. Listed in each figure is the galaxy name, its stellar mass, and its spectroscopic redshift.
The solid line indicates 1 arcsec angular size.}

\end{figure*}

\begin{table*}
 \centering
 \begin{minipage}{140mm}
  \caption{General properties of the selected galaxies}
  \begin{tabular}{ccccccc}
  \hline
 Name & R.A.     & Decl.   & Redshift & M$_{abs}$ & Stellar Mass & kpc/arcsec\\
      & (J2000)  & (J2000) &          & (B-band)  & ($\times$10$^{10}$ M$_{\sun}$) & \\
  \hline
 UDF3372 & 03 32 42.3  & -27 47 46     & 0.996 & -21.7 &  2.4 & 8.001 \\
 UDF5417 & 03 32 39.9  & -27 47 15     & 1.095 & -21.6 &  3.3 & 8.166 \\
\hline
\end{tabular}
\end{minipage}
\end{table*}

\subsection{Profile Extraction}

We have extracted the surface brightness profiles in all the bands available for each of the galaxies. In the case of UDF5417, we have
obtained both WFC3 and NICMOS profiles, but we only show the deepest profiles corresponding to WFC3. The technique used to obtain the
surface brightness profiles is fully explained in Bakos \& Trujillo (2012). Summarizing, we extract radial surface brightness profiles
on masked images in order to avoid contamination on our light profiles. We apply conservative masking on to sources which clearly do not
belong to the galaxy, like foreground stars, background galaxies, etc. These sources are extracted by SExtractor (Bertin \& Arnouts
1996). We use some SExtractor parameters, such as the measured flux, elongation, and similar, to determine the shape and size of these
mask regions. 

In order to extract radial surface brightness profiles representative of the most external part of the galaxy, we need to get
characteristic values of the ellipticity and position angle of this region. We have done this by computing the second-order moment of
the light distribution of the galaxy using the z-band image. The second-order moment is directly related to the position angle, the
semimajor (A) and semiminor (B) axis lengths. We fix this ellipticity and position angle for all elliptical apertures. In each aperture, we estimate the galaxy flux by the 3$\sigma$ rejected mean of the
pixel values of that aperture. This helps to minimize the effect of morphological features like a spiral arm crossing the aperture. 

Although our images are sky substracted, our light profiles, however, can still be contaminated by some residual local sky background.
This local background is estimated by using equally spaced apertures. We obtain the number count profiles of the galaxy up to very far
distances, and we chose a large aperture where the profiles become flat, beyond the identifiable profiles of the galaxy. After a
careful analysis of the profiles in all bands, the region where the sky is determined was chosen to be located at 55$<$R(kpc)$<$75 kpc
(see Fig. \ref{masks}). The 3$\sigma$ rejected mean of the fluxes inside this aperture gives a robust estimate of the residual local
sky background (for further information, see Pohlen \& Trujillo 2006). The error of the background determination is key to decide the
surface brightness level down to which we trust our profiles. Following a conservative approach,this is placed where the profiles
obtained by either over- or undersubtracting the sky measurement by $\pm$$\sigma$ start to deviate with more than 0.2 mag from the
original profiles.

\begin{figure*}
\includegraphics[width=18cm]{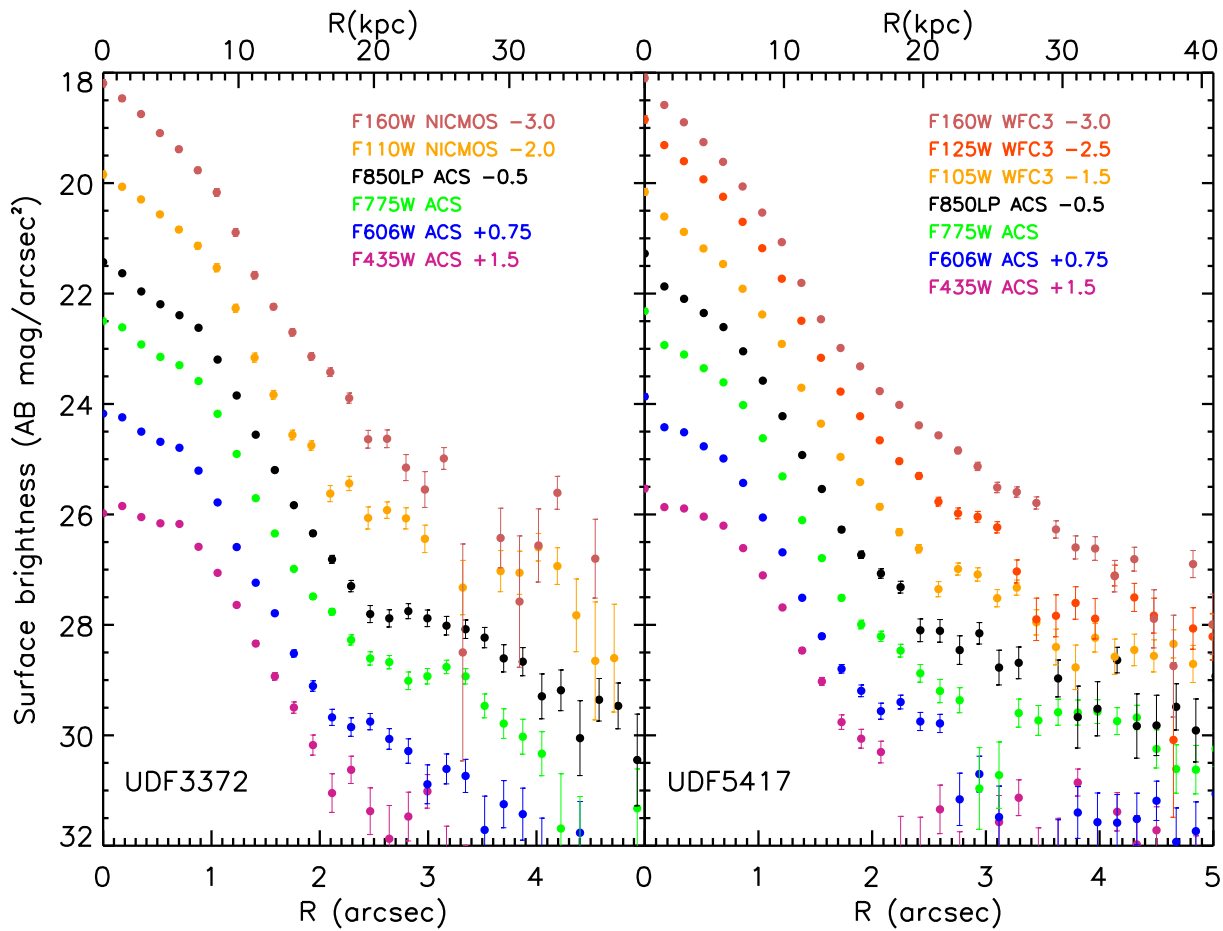}

 \caption{Observed profiles of the UDF3372 (ACS and NICMOS) and UDF5417 (ACS and WFC3) galaxies. The
break features in the discs of the galaxies at  $\sim$1 arcsec (i.e. $\sim$8 kpc) are observed in all the
bands. Beyond R$=$2 arcsec, in all the cases, there is an excess of light over the outer disc
(1$<$R$<$2 arcsec) light exponential decrement. We identify this excess as the growing effect of the
stellar halo in the surface brightness profiles of the galaxies.} 

\label{observedprofiles} \end{figure*}

The observed profiles are shown in Fig. \ref{observedprofiles}. Different vertical offsets have been
applied to the profiles to facilitate the comparison among the different bands.

\section{analysis}

\subsection{The shape of the surface brightness profiles}

The two most remarkable features shown in the surface brightness profiles of Fig. \ref{observedprofiles} are the presence of a break in the disc at
around 1 arcsec (i.e. $\sim$8 kpc at z=1), and some extra light above the outer disc expectations starting at R$\sim$2 arcsec. These
two features are observed in all the bands and in the two galaxies. This strongly indicates that these features are real and are not
connected to sky subtraction problems. We also note that they are observed in the NICMOS data (which are shallower than the WFC3),
pointing out that (although the profiles are noisier) these data are still deep enough to detect the excess of light in the outer
regions of the galaxies. We have checked that this in fact the case using the NICMOS F160W filter of UDF5417 and comparing this to the
WFC3 F160W of the same object. The excess of light is also apparent in the NICMOS profile. The position of the break in the disc (at $\sim$1 arcsec) is pretty stable, independently of the different spatial resolution of the different bands used. We note, however, that in the very
inner region of the UDF3372, the presence of a bulge could be masked by the worst resolution of the NICMOS imaging compared to the WFC3
data. 

We work in this paper under the hypothesis that the excess of light at R$>$2
arcsec (i.e. R$>$16 kpc) is due to the presence of stellar haloes in these
objects. Our hypothesis is based on the similar shape of the surface brightness
distribution of our galaxies compared to those found in nearby disc galaxies. In
the local Universe, both based on integrated photometry (i.e. Bakos \& Trujillo
2012) as well as using stellar counts (see e.g. Barker et al. 2012), the outer
regions of low-inclined disc galaxies (R$>$10 kpc) are characterized by an
exponential surface brightness decrement followed by an excess of light over
this exponential decay. This very outer excess of light is located at
R$\gtrsim$20 kpc and has surface brightness $\mu$$_R$$>$28 mag/arcsec$^2$. These
faint surface brightness are equivalent  to the ones found in the literature
using galaxies with edge-on orientations at similar radial distances above the
galactic planes. For instance, the edge-on spiral NGC4565 has, along its minor
axis, a surface brightness of $\mu$$_{6660\AA}$=27.5 mag/arcsec$^2$  (i.e.
$\mu$$_R$$\sim$28 mag/arcsec$^2$ in R-band) at 22 kpc above the disc (Wu et al.
2002). Similarly,  Jablonka et al. (2010) also found that the stellar halo of
the edge-on disc NGC3957 has surface brightness of 28.5 mag/arcsec$^2$ in R-band
at 20 kpc over the disc plane. We conclude, consequently, that at distances of
20 kpc and surface brightness of $\mu$$_R$$>$28 mag/arcsec$^2$ both in face-on
and edge-on orientation of nearby disc galaxies we are observing the same
component of the galaxy, i.e. their stellar haloes. By analogy, we consider that
the  excess of light over the exponential disc decrement located in the
peripheries (i.e. R$>$20 kpc) of our high-z face-on disc galaxies can be
attributed to the stellar halo of these objects.

However, we should note, before claiming that the excess of light is our high-z
galaxies are only due to their stellar haloes, that de Jong (2008) showed that the effect of
very extended PSF tails on the measurements of halo
light can be significant  (especially in the case of edge-on galaxies). de Jong
conducted this study both in the Sloan Digital Sky Survey (SDSS) and in the HUDF.
Although de Jong (2008) concluded that in face-on orientation the effect of the
PSF should not be so significant as in the edge-on case, it is worth exploring
how much this effect  can alter the estimation of the amount of light in the
outer regions of our high-z galaxies. We explore this issue in the next subsection.

\subsection{The effect of the PSF on the surface brightness profiles}

To estimate the contribution from scattered light at large radii in our surface
brightness profiles is necessary to model the intrinsic light distribution of the
galaxies. We have done this by assuming that the galaxies are well described by
a S\'ersic r$^{1/n}$ (1968) bulge plus a double-exponential disc. This
accurately describes the inner ($\lesssim$2 arcsec) profile. To calculate the
observed distribution from the model distribution, the PSF needs to be determined
out to a large radius ($\sim$10 arcsec; de Jong 2008). Unfortunately, such an
extended PSF cannot be accurately measured from the Hubble UDF images itself as
there are too few bright stars. Following de Jong (2008), we have therefore used
the TINYTIM HST PSF modelling software to create artificial PSFs for each of our
bands. 

\begin{figure*}
\includegraphics[width=\textwidth]{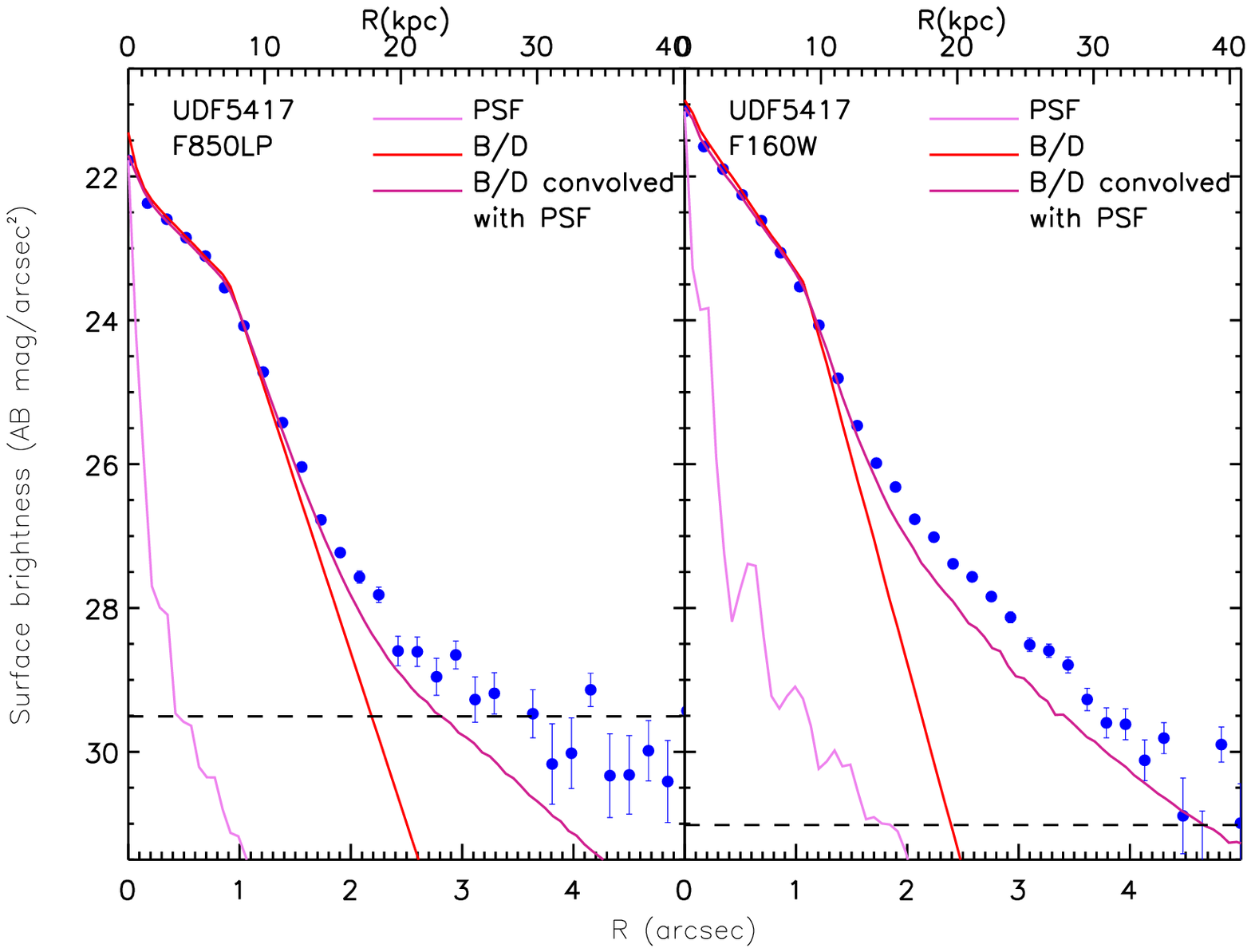}

\caption{Surface brightness profile of the UDF5417 galaxy in the ACS F850LP (left-hand side) and the
WFC3 F160W (right-hand side) bands. The PSF profiles normalized to the same central surface brightness
of the galaxy, as well as the galaxy models plus the galaxy models convolved with the PSFs are shown.
The dashed lines in both panels indicate the limiting surface brightness down to which the surface
brightness profiles are reliable.} 

\label{psfeffect5417detail} 
\end{figure*}

To illustrate the effect of the PSF, we show in Fig. \ref{psfeffect5417detail} the model galaxy for UDF5417 convolved with the
PSF for the two reddest bands in both the ACS and in the WFC3 camera. For the
rest of the bands and for galaxy UDF3372, the PSF effect is shown in the
Appendix. Clearly, a large fraction of the light seen in the surface brightness
profile above the galaxy model is due to scattered light from the extended PSF.
However, there is still a clear excess of light beyond 2 arcsec that can not
be explained by the PSF effect alone. It is necessary to add an extra component in the outer
region of the galaxy to explain this excess of light.

\begin{figure*}
\includegraphics[width=12cm]{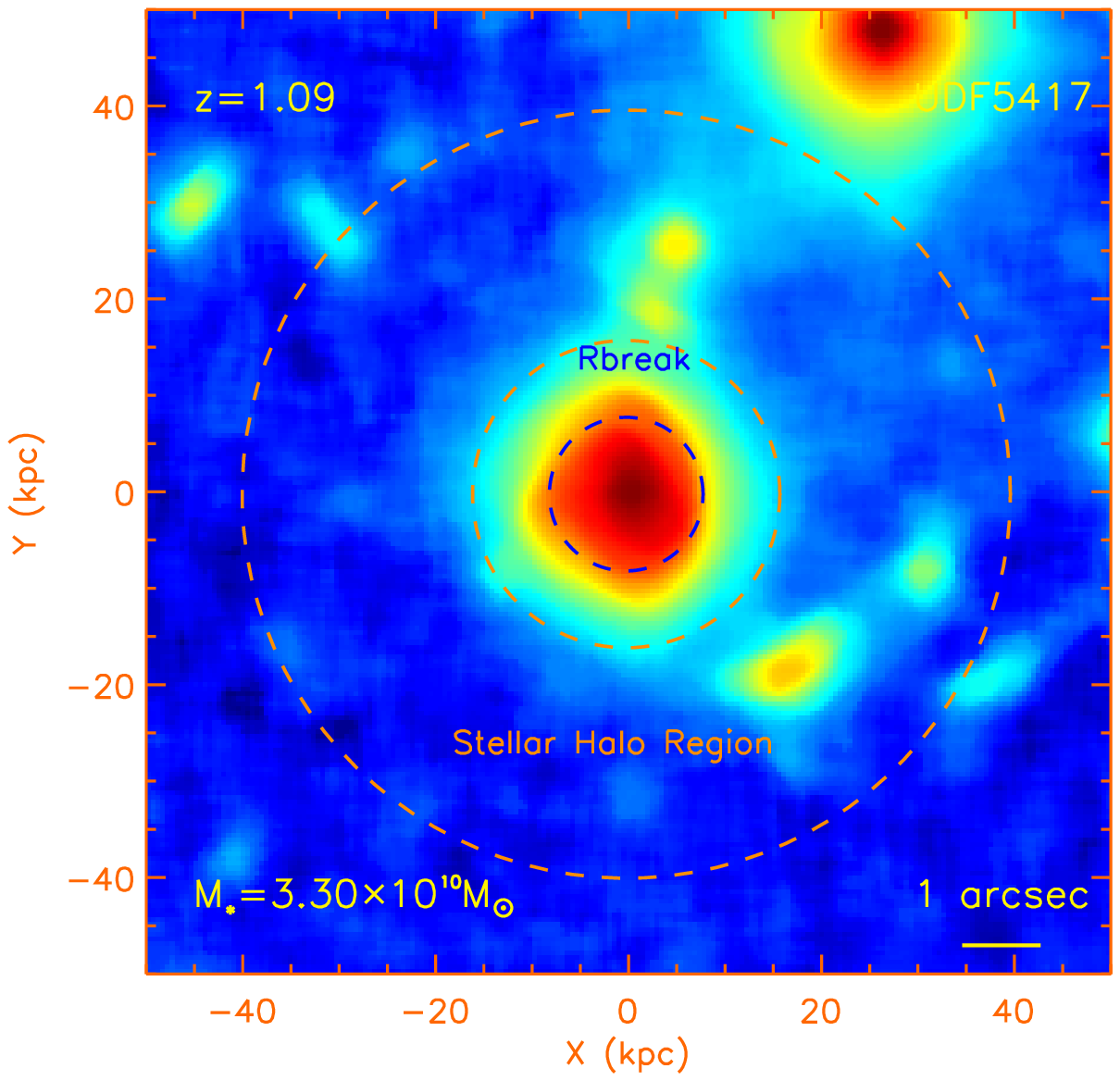}

\caption{Stacked image of all the WFC3 bands (F105W, F125W and F160W) of the UDF5417 (z=1.09) galaxy. The image has been filtered with
a median filter of 5 pixels width. These allow us to see the faintest features of the object. The dashed blue circle indicates the
position of the break on the surface brightness profile of the disc of the galaxy. The region enclosed by the two orange dashed circles
corresponds to the region of the stellar halo explored in this paper. The visible structure of the halo is not perfectly symmetric
suggesting some on-going accretion.} 

\label{halo} \end{figure*}

In order to further investigate which is the origin of the outer light,  we show in Fig. \ref{halo}
the result of stacking the images from the three WFC3 bands of UDF5417. This stacked image has been
filtered with a median filter of 5 pixels width to enhance the lowest surface brightness features of
the galaxy. The region corresponding to the excess of light above the outer disc expectations observed
in the profiles (i.e. 16$<$R$<$40 kpc) are enclosed by two orange dashed circles. We can ascertain
that this region is well beyond the main body (disc) of the galaxy. We also observe that the
distribution of light in that region is not perfectly symmetrical. This absence of symmetry in a region
beyond the disc could be an indication of ongoing accretion.

In the local Universe, deep surface brightness profiles (10 magnitudes in range) of spiral galaxies,
as the ones shown here for z=1 objects, also exhibit the excess of light beyond the outer disc (see
e.g. Bakos \& Trujillo 2012). In present-day spirals, the surface brightness ($\mu_r$$\gtrsim$28
mag/arcsec$^2$), colors (g-r$\gtrsim$0.6) as well as the radial distances (R$\gtrsim$15 kpc) of these
light excesses correspond to the expected characteristics of stellar haloes.

\subsection{Bulge-disc-halo decomposition}

\begin{figure*}
\includegraphics[width=\textwidth, angle=0]{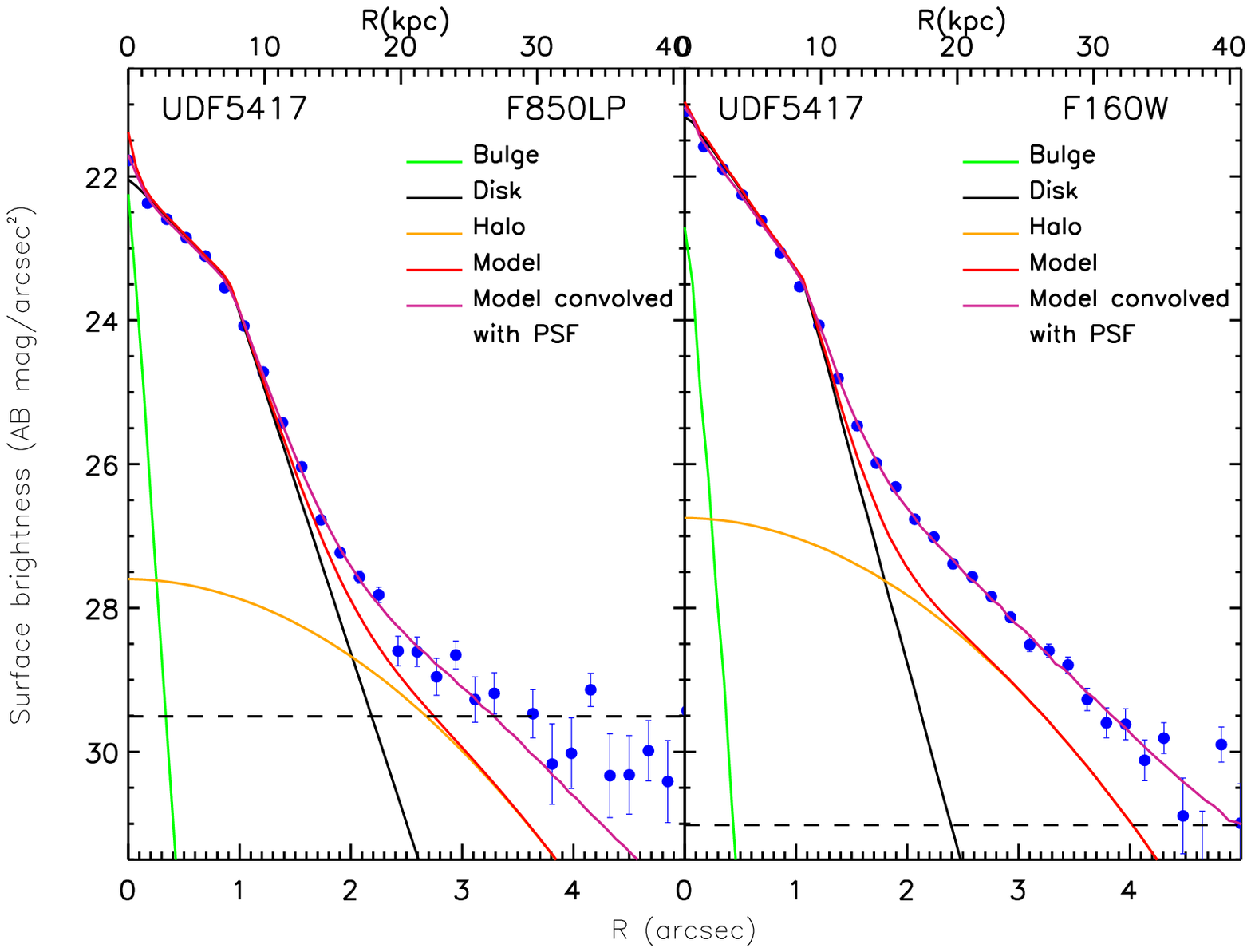}

 \caption{Bulge-disc-stellar halo decomposition of the galaxy UDF5417 in the
reddest bands of the ACS (F850LP) and the WFC3 (F160W). The bulge, disc, halo as
well as the galaxy models plus the galaxy models convolved with the PSFs are
shown. The dashed lines in both panels indicate the limiting surface brightness
down to which the surface brightness profiles are reliable.}

\label{decompositions}
\end{figure*}

As we have seen in the previous section, the effect of the PSF is relevant in the outer regions of our
galaxies and affects differently depending on the observed wavelength. To account for this effect, we
have carried out a bulge-disc-stellar halo decomposition of our galaxies in all the observed
bands\footnote{Note that in the case of UDF3372, it was not necessary to include a bulge to
the decomposition as there was not evidence of any surface brightness increase in the centre of the
profiles. To get our analysis as simple as possible, we just conducted a disc-stellar halo
decomposition for this object.}. Those models allow us to work with the galaxies profiles without being
affected by the PSF. From those PSF deconvolved models, we will obtain (see next section) the restframe
surface brightness profiles and the colour profiles of our objects not affected by the
PSF. As we have done before, for illustrative purpose, we show the bulge-disc-stellar
halo decomposition for the galaxy UDF5417 in the F850LP and F160W WFC3 bands (Fig.
\ref{decompositions}) and leave the rest of the bands for the Appendix.

The structural parameters of the bulge-disc-stellar halo decomposition for each band are shown in
Table \ref{bulgedischalo}. The structural decomposition is conducted on the 1D (i.e. 2D
azimuthally averaged) profiles using our own code. Due to the large number of structural parameters
involved in this decomposition, we have fixed several parameters. For instance, the size and shape of
the bulge are based on the information retrieved from the F850LP band. In this band, the combination
of depth and resolution for this structural component is the best that we have in our data. The
magnitude of the bulge in the other wavelengths is then left free during the fitting process.
Similarly, the shape (Gaussian) and size of the halo are also fixed and only the brightness is left
free during the fit. In the case of the stellar halo, the choice of the shape and size are based on
both the appearance of the surface brightness in multiple bands (see Appendix) as well as in the
information retrieved from local galaxies for their stellar haloes (see e.g. Bakos \& Trujillo 2012).
Nonetheless, other stellar haloes models have been used in the literature. For example, in the case of
the halo of M31, exponential profiles for describing the surface brightness of this component have
been tried by Irwin et al. (2005), Ibata et al. (2007), Tanaka et al. (2010) or Gilbert et al. (2012).
We have also fitted our stellar haloes using exponential profiles. We discuss how this affects  the
structural parameters of the haloes and the results of our paper in the Appendix.  In what follows, we
explore the restframe properties (surface brightness and colours) of our disc galaxies at z$\sim$1.

\begin{table*}
 \centering
 \begin{minipage}{140mm}
  \caption{Bulge-disc-halo decomposition of the selected galaxies}
  \begin{tabular}{cccccccccc}
  \hline
  & & & & & UDF3372 & & & & \\ 
  \hline
 Band & m$_{bulge}$  & r$_{e,bulge}$     & n$_{bulge}$   &  m$_{disc}$ & h$_{i,disc}$  & r$_{break}$ & h$_{o,disc}$ & m$_{halo}$ & r$_{e,halo}$\\
      &        & (arcsec)  &     & &   (arcsec)        &  (arcsec) & (arcsec)                 &     & (arcsec)     \\
  \hline
b     & ... & ... & ... & 23.12  & 1.20 & 0.90 & 0.30 &  26.30 & 2.5 \\
v     & ... & ... & ... & 22.45  & 0.95 & 0.90 & 0.24 &  25.70 & 2.5 \\ 
i     & ... & ... & ... & 21.57  & 0.85 & 0.97 & 0.22 &  25.00 & 2.5 \\
z     & ... & ... & ... & 21.10  & 0.75 & 1.00 & 0.24 &  24.60 & 2.5 \\
F110W & ... & ... & ... & 20.76  & 0.70 & 1.15 & 0.17 &  24.40 & 2.5 \\ 
F160W & ... & ... & ... & 20.25  & 0.50 & 1.17 & 0.18 &  24.20 & 2.5 \\
 \hline
  & & & & & UDF5417 & & & & \\ 
      \hline
b     & 28.6 & 0.08 & 1 &  23.10 &  1.25 & 0.80 & 0.30 & 26.90 & 1.7  \\ 
v     & 27.1 & 0.08 & 1 &  22.68 &  0.95 & 0.85 & 0.22 & 25.60 & 1.7  \\ 
i     & 26.6 & 0.08 & 1 &  22.01 &  0.70 & 0.90 & 0.22 & 25.60 & 1.7  \\ 
z     & 25.7 & 0.08 & 1 &  21.45 &  0.68 & 0.94 & 0.23 & 24.80 & 1.7  \\ 
F105W & 25.5 & 0.08 & 1 &  21.30 &  0.55 & 0.99 & 0.22 & 24.60 & 1.7  \\ 
F125W & 25.3 & 0.08 & 1 &  21.04 &  0.50 & 1.05 & 0.20 & 24.70 & 1.7  \\ 
F160W & 25.0 & 0.08 & 1 &  20.85 &  0.48 & 1.10 & 0.19 & 23.95 & 1.7  \\ 
\hline
\end{tabular}
\label{bulgedischalo}
\end{minipage}
\end{table*}

\subsection{Restframe surface brightness and colour profiles}

\begin{figure*}
\includegraphics[width=18cm]{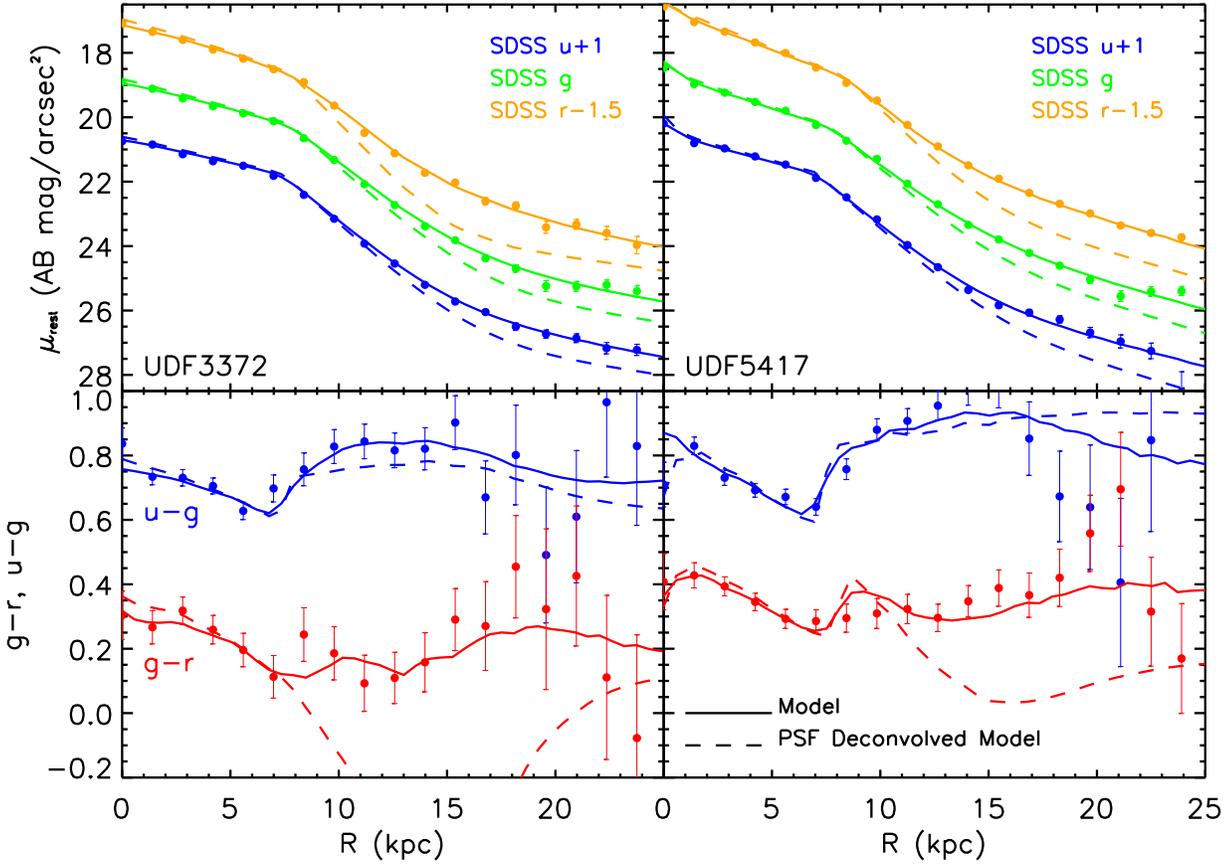}

 \caption{{\it Upper panels:} the u, g and r-band equivalent rest-frame surface brightness profiles obtained by the interpolation of the
 observed band profiles.  The solid lines are the rest-frame surface brightness profiles obtained by the interpolation of the model fits to the observed band profiles (see Fig. \ref{decompositions}; violet line). The dashed lines are the rest-frame surface brightness profiles obtained by the interpolation of the PSF deconvolved model fits to the observed bands (see Fig. \ref{decompositions}; red line). The rest-frame surface brightness profiles have been corrected by the cosmological dimming. {\it Bottom panels:} The g-r and u-g rest-frame color profiles are shown. The solid and dashed lines are the colours obtained from the model fits to the data (solid line) and the PSF deconvolved models (dashed lines).}
\label{restprof}
\end{figure*}

Using the observed surface brightness profiles, we have built the SEDs at different radial distances.
We have linearly interpolated among the different observed filters to obtain the flux in the
equivalent SDSS restframe bands: u, g and r. After doing this, we have corrected the rest-frame surface
brightness profiles by adding the equivalent magnitude to the cosmological dimming at those redshifts.
We show the rest-frame surface brightness profiles of our galaxies in Fig. \ref{restprof}. The
rest-frame surface brightness profiles are used to build the rest-frame radial color profiles as well. 

The light in the outer regions of our galaxies is contaminated by the light scattered from the inner
parts of the galaxy due to the extended PSF. The level of contamination is different depending on the
wavelength. To account for that, both in the rest-frame surface brightness profiles as well as in the
rest-frame color profiles, we have used the bulge-disc-halo fit described in the above section. As we
have done with the observed profiles, we have linearly interpolated among the different galaxy models
of each observed filter to obtain the surface brightness of the galaxy model in the equivalent SDSS
rest-frame band. The result of doing this is shown in Fig. \ref{restprof}. As we can see the models
reproduce fairly well both the rest-frame profiles and the rest-frame colours. Now, we repeat the
exercise but this time using the PSF deconvolved models of each observed band (i.e. the model fits but
without convolving them with the PSF). The new ("PSF effect free") restframe surface brightness and
color profiles are shown in Fig. \ref{restprof} with dashed lines.

At it is expected, the outer regions of the PSF deconvolved rest-frame surface brightness profiles are
fainter than when the PSF effect is not accounted for. The  effect on the u-g rest-frame colour is not
very significant as this colour is mostly based on the observed ACS bands (which present the smaller
PSFs). In the case of the g-r rest-frame colour, the change is larger since this colour is obtained from
the information provided  by the NIR cameras\footnote{We warn the reader about the dip
present in the deconvolved (g-r) color between 10 to 20 kpc in the case of UDF3372. We think this dip
is artificial and not a real reflect of the underlying light distribution of the galaxy. This dip
corresponds to the disc-stellar halo transition region and it is mainly produced by the shape of the
deconvolved r-band profile for this galaxy. The rest-frame r-band mostly reflects the shape of the
observed F110W band. Contrary to UDF5417, where we are using much better resolution NIR data, in the
case of UDF3372 this radial region is strongly affected by the large wing of the NICMOS PSF (see e.g.
Figure B1) and consequently, the results in this region are very uncertain.}.

From Figure \ref{restprof}, it follows immediately too that the surface brightness of our high-z
galaxies are brighter ($\sim$2 mag/arcsec$^2$) than those shown in locally equivalent stellar mass
galaxies. This is expected due to the large star formation activity and younger stellar population of
those objects $\sim$8 Gyr ago. Fortunately for us, not only the disc of the galaxies are brighter but
also the stellar haloes. This is the reason why we can observe them even considering that the surface
brightness of these galaxies are severely affected by the cosmological surface brightness dimming  at
z$\sim$1 ($\sim$3 mag/arcsec$^2$).  The radial colour profiles of our galaxies (more clear in the data
itself than in the models) show the U-shape (associated to the break in the stellar disc) which has
been observed in the nearby Universe (Bakos et al. 2008) as well as in z$\lesssim$1 disc galaxies
(Azzollini et al. 2008a). In agreement with the fact that high-z galaxies have younger stellar
populations at z=1 than now, the observed colours of our disc galaxies are bluer than in local
objects. For instance, whereas in the local Universe, the surface brightness profiles usually have  a
(g-r)=0.47$\pm$0.02 mag at the break radius, our galaxies present (g-r)$<$0.3  mag at this feature.

In Table 3, we provide some structural parameters of the disc component of the galaxies. In particular, we present the radial position of
the disc break and the r-band restframe surface brightness  (after correcting by the cosmological dimming) at this position. Both the values
that we get for the surface brightness of the breaks ($\mu_{r,break}$$\sim$20.2 mag/arcsec$^2$) and their radial positions
(R$_{break}$$\sim$8.1 kpc) are typical of the population of disc galaxies with similar absolute magnitudes and stellar mass at those
redshifts (e.g. Azzollini et al 2008b). In the next section, we concentrate on the properties of the stellar haloes.

\begin{table*}
 \centering
 \begin{minipage}{140mm}
  \caption{Structural properties of the selected galaxies in the restframe SDSS r band.}
  \begin{tabular}{cccccccc}
  \hline
 Name & R$_{break}$  & R$_{break}$     & $\mu_{r,break}$   &  R$_{e,halo}$ & $<\mu_{e,halo}>$  & L$_{halo}$/L$_{total}$ & (g-r)$_{halo}$ \\
      &  (arcsec)       & (kpc)  & (mag/arcsec$^2$)    &  (kpc)         &  (mag/arcsec$^2$) & (r-band)                &  (mag)           \\
  \hline
 UDF3372 &   0.95$\pm$0.02        &  7.6$\pm$0.1    &      20.0$\pm$0.03             &   19.6$\pm$2.3          &   25.4$\pm$0.25  &   0.040$\pm$0.010    & 0.05$\pm$0.30 \\
 UDF5417 &   1.05$\pm$0.02        &  8.6$\pm$0.1    &      20.3$\pm$0.03             &   13.6$\pm$0.6          &   24.4$\pm$0.10  &   0.041$\pm$0.004    & 0.15$\pm$0.15 \\
\hline
\end{tabular}
\end{minipage}
\label{structure}
\end{table*}

\subsection{Stellar halo properties}

\begin{figure*}
\includegraphics[width=\textwidth, angle=0]{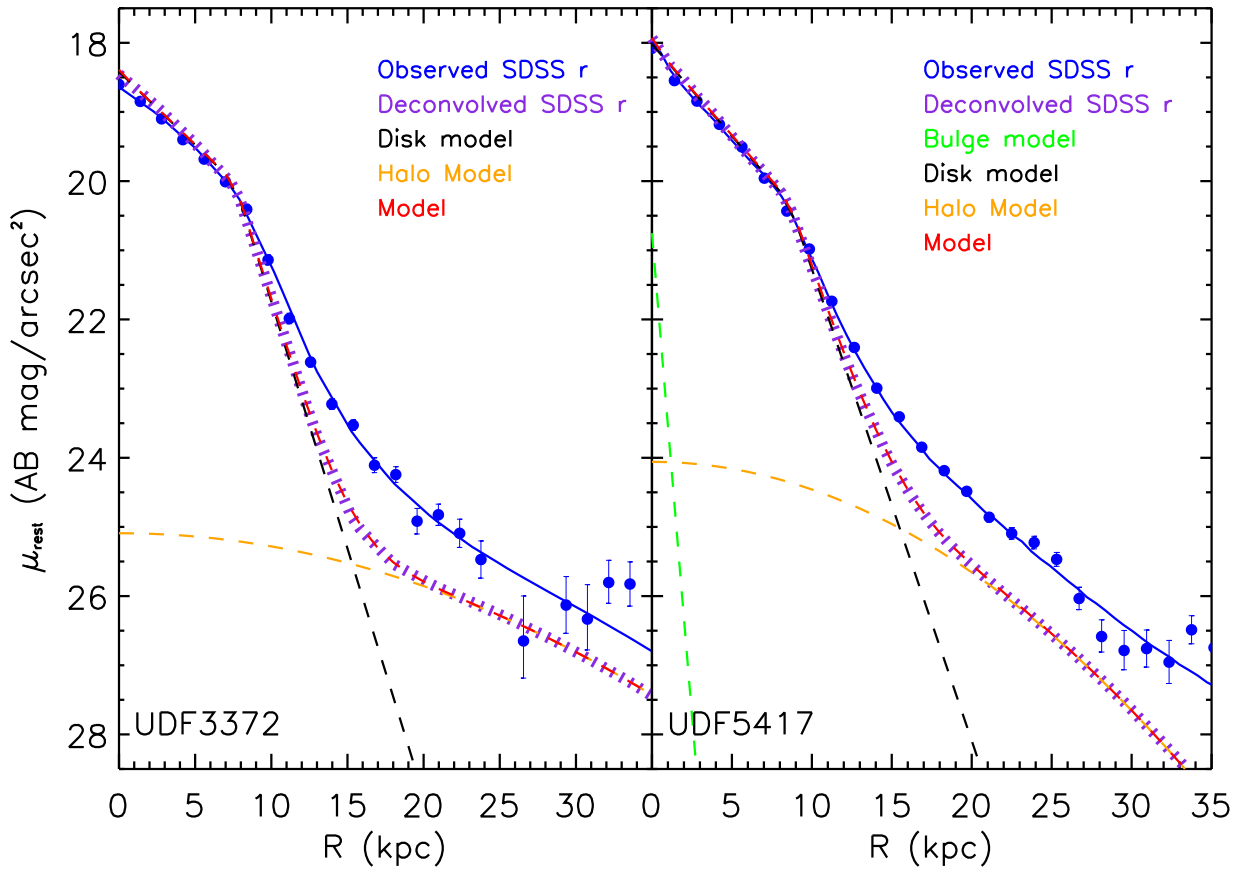}

 \caption{Bulge-disc-stellar halo decomposition of the PSF deconvolved r-band rest-frame surface brightness profiles of UDF3372 and UDF5417. The
 profiles have been corrected by the cosmological dimming. }
\label{decomposition}
\end{figure*}

Following a similar analysis to the one presented in Bakos \& Trujillo (2012) for local galaxies and
as we have done earlier, we characterise the structural properties of the stellar haloes at high-z by
decomposing the PSF deconvolved r-band surface brightness rest-frame profile into three different
components: a S\'ersic r$^{1/n}$ (1968) bulge, a double-exponential disc, and a stellar halo described
by a S\'ersic model with n=0.5 (see Fig. \ref{decomposition}). This later choice seems somehow
arbitrary but it has been chosen because it represents well the shape of the profile in the outer
region. It also allows a direct comparison with the analysis of the stellar haloes
conducted at z=0.

In Table 3, we provide the effective radii of the haloes and their mean surface
brightness according to the rest-frame r-band. In the same table, we also show the
fraction of light that is contained in the stellar haloes compared to the total
light of the galaxies (L$_{halo}$/L$_{total}$). The effective radii of our
haloes (R$_{e,halo}$$\sim$17 kpc) are within the range of the ones found in the
local Universe. The most important difference of the high-z stellar haloes
compared to the local ones is in connection with their mean effective surface
brightness. At high-z, the values found $<\mu_{e,halo}>$$\sim$24.9
mag/arcsec$^2$ (r-band) are significantly much brighter ($\gtrsim$ 3
mag/arcsec$^2$) than those measured in nearby galaxies. We finally have obtained
the (g-r) colour at the stellar halo position by using the PSF deconvolved model
color profiles  at 20$<$R$<$25 kpc (see Fig. \ref{restprof}). Ideally, we would
like to explore these colours farther away but the errors are so large that this
estimation becomes prohibitive. For this reason, we concentrate on the above
region. Nonetheless, if we trust the bulge-disc-halo decomposition conducted
here, these colours should not be significantly affected by the colours of the
outer disc population after the PSF effect correction. To be conservative, the
halo colour error  as well as the error on the mean effective surface brightness
of the halo is based on the observed error measured in the surface brightness 
and colour distributions at the farthest halo regions we can explore with
confidence (i.e. 20$<$R$<$25 kpc). We acknowledge that these errors do not
account for potential variations of the shape of the halo (which we have fixed
here to a Gaussian n=0.5 shape). So, we warn the reader that there is a
potential source of systematic bias that cannot be easily quantified without having
deeper and/or higher resolution images.

\subsection{Comparison with the local sample}

\begin{figure*}
\includegraphics[width=\textwidth]{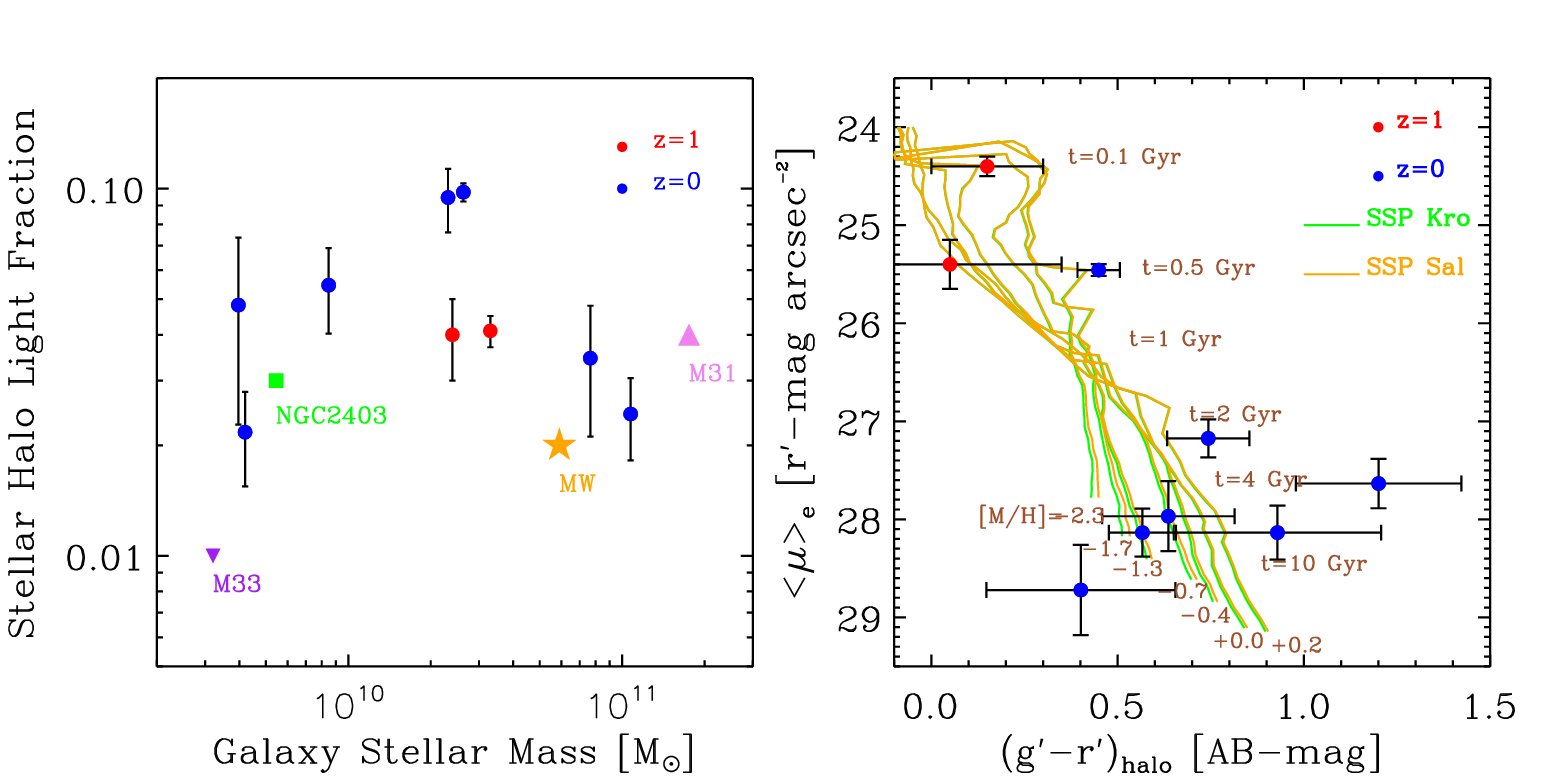}

 \caption{{\it Left-hand panel}: fraction of the stellar halo light in the r-band (L$_{halo}$/L$_{total}$) versus the total galaxy mass. The stellar halo light is
estimated by fitting  a n=0.5 S\'ersic-model to the region of the stellar halo of the galaxies; meanwhile, to obtain the total galaxy
light we integrated the observed profile. The red points are the galaxies at z$\sim$1  and the blue points a reference local sample
(Bakos \& Trujillo 2012). The stellar halo light fraction observed in the Milky Way (Carollo et al. 2010), M31 (Courteau et al. 2011),
M33 (McConnachie et al. 2010) and NGC2403 (Barker et al. 2012) are also overplotted. The light contribution of the stellar halo to the
total galaxy light varies from 1 to 5 per cent, but in case of ongoing mergers (the two most extreme cases in the local reference sample)
the stellar halo light fraction can be as high as 10 per cent. {\it Right-hand panel}: mean effective surface brightness of the stellar halo versus
the (g-r) colour of the stellar halo. Overplotted are the colour and surface brightness tracks predictions from the
Vazdekis et al. (2010) models for SSPs with two different IMFs (Salpeter 1955 and Kroupa 2001).}

\label{comparison}
\end{figure*}

To put our high-z galaxies in context, we have compared the properties of their stellar haloes with the ones found in
the local Universe. The result of this comparison is presented in Fig. \ref{comparison}. In the left-hand panel, we have
the light fraction of the stellar haloes (in the restframe SDSS r-band) versus the total galaxy stellar mass. We have
taken the local values from different authors. These works are listed in the caption of Fig. \ref{comparison}. We find
that the amount of light contained in the high-z stellar haloes  is very similar to the values found in our local
reference sample ($\sim$4\%). This result implies that, whatever the complex mechanisms acting on the evolution of the
galaxies as a whole and the stellar haloes as a separate component, they should produce a decrease of the brightness in
both structures which is similar in order to preserve such fraction constant with time.

In the right panel of Fig. \ref{comparison}, we compare the mean effective surface brightness and the (g-r) color of
the stellar haloes both at high and low-z. As we have been claiming through the paper, the stellar haloes at high-z
are significantly much brighter than today. Also, their colours are bluer ((g-r)$\sim$0.1 mag) than in
the local universe ((g-r)$\sim$0.7 mag; see also Zibetti et al. 2004; Zackrisson et al. 2006; Bergvall et
al. 2010). Can the high-z stellar haloes in the high-z Universe evolve into the ones that we
observe locally by passive evolution? The most simple exercise that we can conduct to address this question is to
follow the tracks of single stellar populations (SSPs) with cosmic time and to probe whether the passive evolution of
the stellar population in the haloes can reproduce the local properties. These tracks have been constructed using  the
Vazdekis et al. (2010) models for SSPs with two different IMFs (Salpeter 1955 and Kroupa 2001). We find that the
passive evolution moves the colour and the surface brightness of the high-z stellar haloes towards the observed values
in the local Universe. We explore the consequences of this finding in the next section.

\section{Discussion and Conclusions}

In the current paradigm, stellar haloes of galaxies have a dual nature. The outer halo formed primarily by accretion/mergers, whereas
the inner halo is formed mainly through a dissipative collapse (as a result of gas-rich mergers at early times). Recent cosmological
simulations (e.g. Cooper et al. 2010; Font et al. 2011; Tissera et al. 2012)  indicate that both phenomena take place at z$>$1. If this is the case, one
would expect that the stellar haloes will evolve since that epoch by mostly fading without any significant change in their structures. Our present observations allow us to check this important
prediction of the disc galaxy formation scenario.

According to our data, the structural properties of our high-z stellar haloes 
are compatible with being at place at z$\sim$1. We base this assertion on the
following:  if we assume that the high-z haloes have the same shape than the
local ones (i.e. if they can be well described with n=0.5 profiles as it seems
to be the case), then the sizes of these components (R$_{e,halo}$$\sim$17 kpc)
as well as their contribution to the total light of the galaxies ($\sim$4 per cent) are
very similar to the ones found in the local Universe. The major difference with
respect to the local galaxies is that our haloes are $\sim$3 magnitudes brighter
than their local counterparts. They also present bluer colors than the local
stellar haloes. Both results agree with the idea that high-z stellar haloes are
significantly much younger than present-day ones. According to the SSP
evolutionary tracks that we have explored, the colors of our high-z stellar
haloes correspond to stellar populations with an age $\lesssim$1 Gyr. That
implies that the stars in those haloes were formed basically at 1$<$z$<$2. This
matches very well the theoretical predictions that move most of the formation of
the stellar haloes at those epochs. 

Our sample is so far very small (only 2 objects), but this pilot work can be
easily extended towards other spiral galaxies in the HUDF at lower redshifts and
with different inclinations. Our two disc galaxies have not been pre-selected to
have bright (i.e. detectable) stellar haloes but we admit we cannot infer, with
only two objects, whether this is the general behavior of the whole disc galaxy
population at z$\sim$1. It will be interesting to explore in the future whether
other disc galaxies have prominent stellar haloes as the ones discussed in this
paper. In addition, it will be worth probing whether the properties of other
high-z stellar haloes also fit within a scenario where a pure passive evolution
of their stellar populations matures them towards the present-day Universe stellar haloes.

\section*{Acknowledgments}

We thank the referee for her/his constructive comments that significantly
improved the original manuscript. We are indebted to Pablo G. P\'erez-Gonz\'alez
for allowing us using a private part of the Rainbow database for this work. We
also thank Eric Bell and Patricia Tissera for interesting comments during the
elaboration of this paper. This work has been supported by the Programa Nacional
de Astronom\'ia y Astrof\'isica of the Spanish Ministry of Science and
Innovation under grant AYA2010-21322-C03-02. This work has made use of the
Rainbow Cosmological Surveys Database, which is operated by the Universidad
Complutense de Madrid (UCM).

\bsp

\label{lastpage}

\appendix

\section{Masking}

\begin{figure*}
\includegraphics[width=16cm]{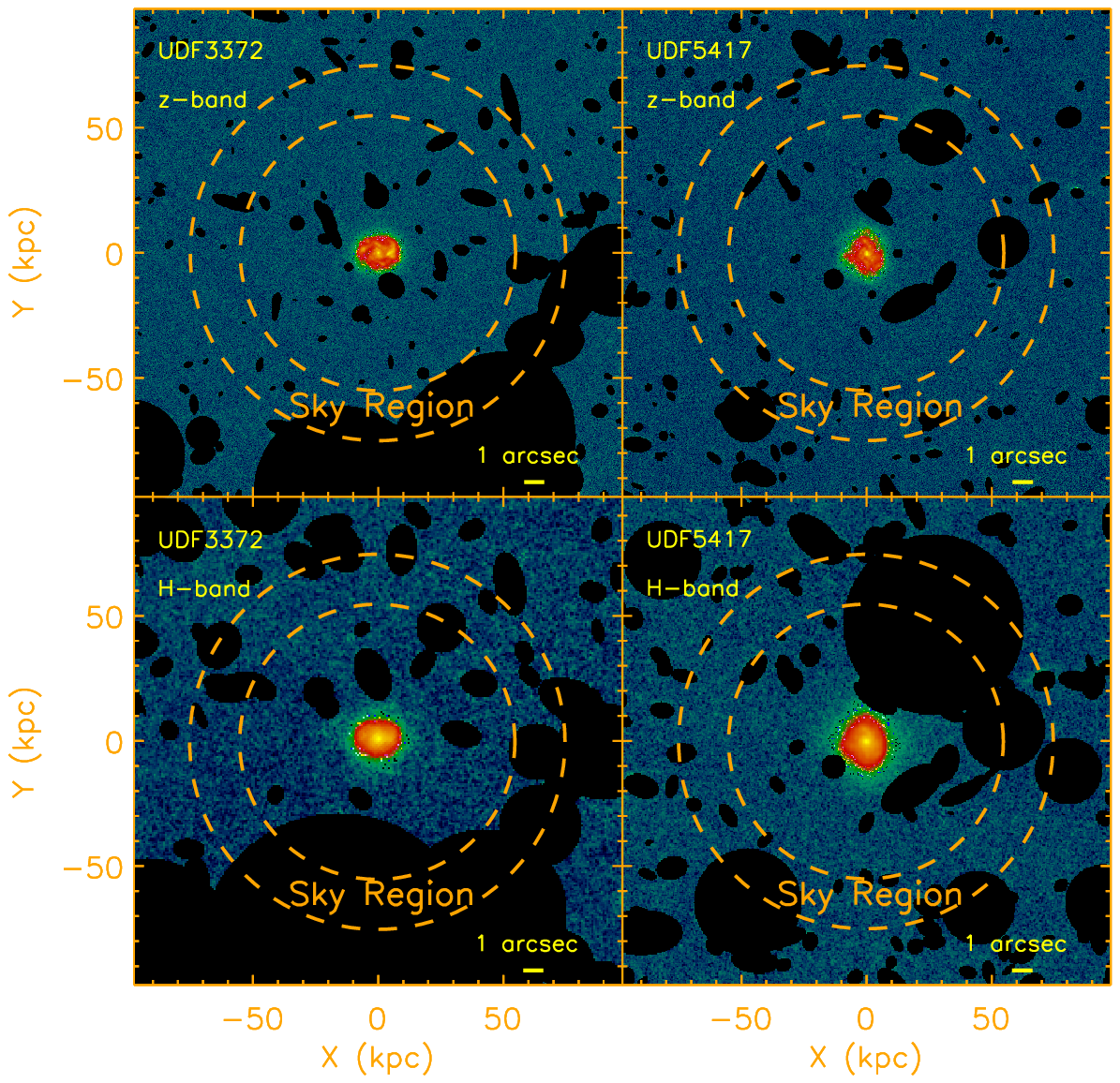}

 \caption{Regions masked (shown in black) in this work. For each regime (optical and NIR), we have
 created different masks to account for the significantly different PSFs and cameras characteristics.
 The region chosen to determine the sky is enclosed by two orange dashed circles.}

\label{masks}
\end{figure*}

The characterization of the stellar haloes (e.g. shape, structural parameters and colours) can be contaminated by the presence of nearby
sources like foreground stars, background galaxies, etc. In this Appendix, we show the masks that we have used in this paper to avoid
this problem (see Fig. \ref{masks}). We created two different mastermasks for each galaxy: one in the optical regime and another one in the NIR. The
background contaminating sources are extracted by SExtractor from a master image composed of all the bands. In the case of the optical
regime, this corresponds to the four ACS filters, and in the case of the NIR data, this master image is composed of  the two NICMOS filters
(for the galaxy UDF3372) and of the three WFC3 filters (for the galaxy UDF5417). The need to separate between the optical and NIR
regimes at creating the masks is justified by the significantly different PSFs sizes and cameras characteristics. Nonetheless, as we
will see later, most of the regions masked in both regimes are in common being the NIR- masked areas larger than in the optical as
expected due to their larger PSFs.

The master images are a stack of the different filters scaled to the z-band flux (optical) and H-band flux (NIR). Sources are
extracted in the "cold mode", optimized for brighter objects. We use some SExtractor parameters such as the measured flux, elongation, and
similar to determine the shape and size of these mask regions.  In some cases, problematic sources, saturated stars, for instance, are
masked manually. Since the masking is based on information coming from all  filters in each regime, we use the resulting masks as 
mastermasks, and it is applied to all filters in the optical regime (mastermask in the optical) and in the NIR (mastermasks in the
NIR).	

\section{PSF effect}

\begin{figure*}
\includegraphics[width=\textwidth]{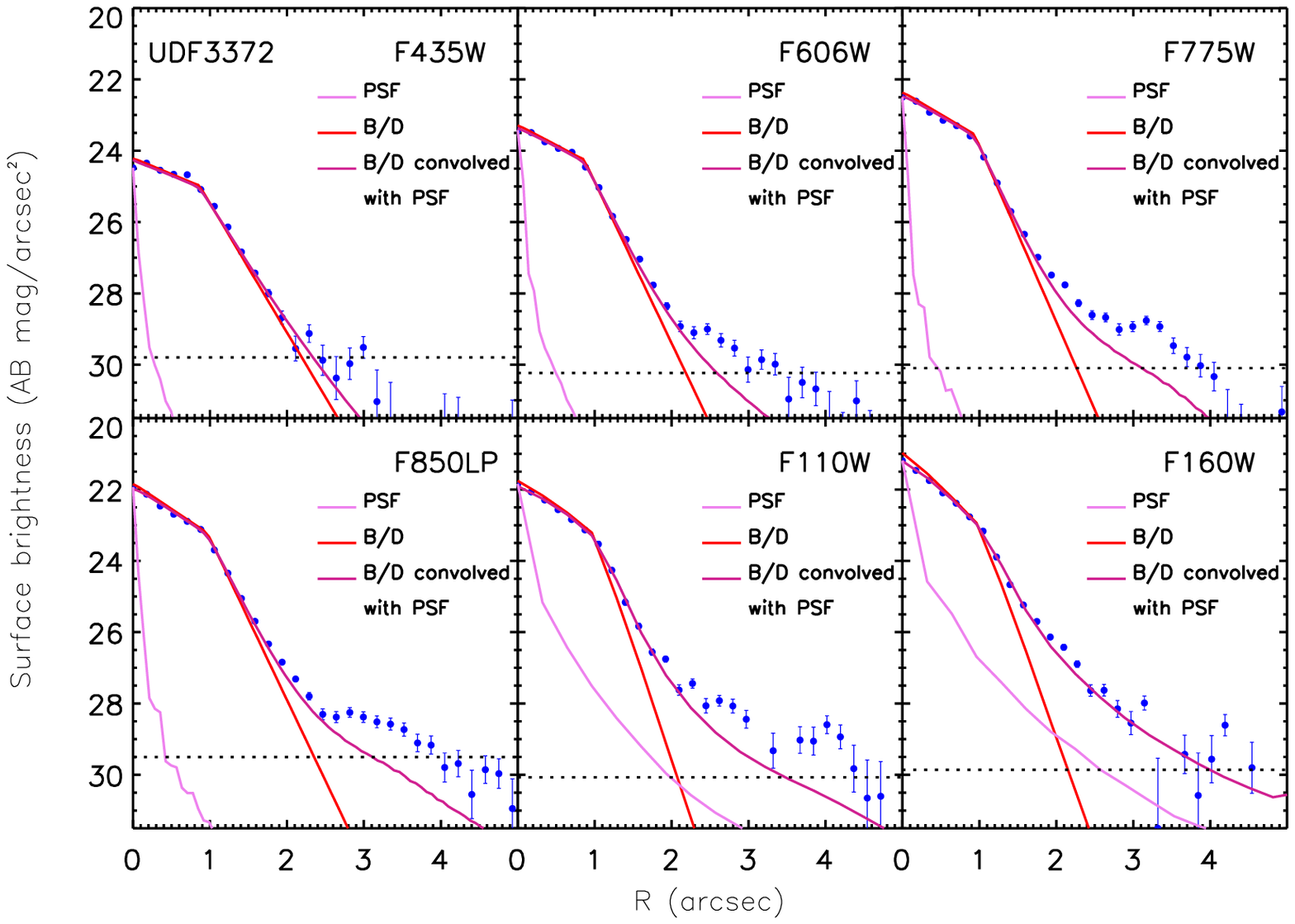}

\caption{Surface brightness profile of the UDF3372 galaxy in all the observed bands. The PSF profiles normalized to the same central surface brightness of the galaxy, as well as the galaxy models plus the galaxy models convolved with the PSFs are shown. The dashed lines in both panels indicate the limiting surface brightness down to which the surface brightness profiles are reliable.} 

\label{psfeffect3372} 
\end{figure*}

\begin{figure*}
\includegraphics[width=\textwidth]{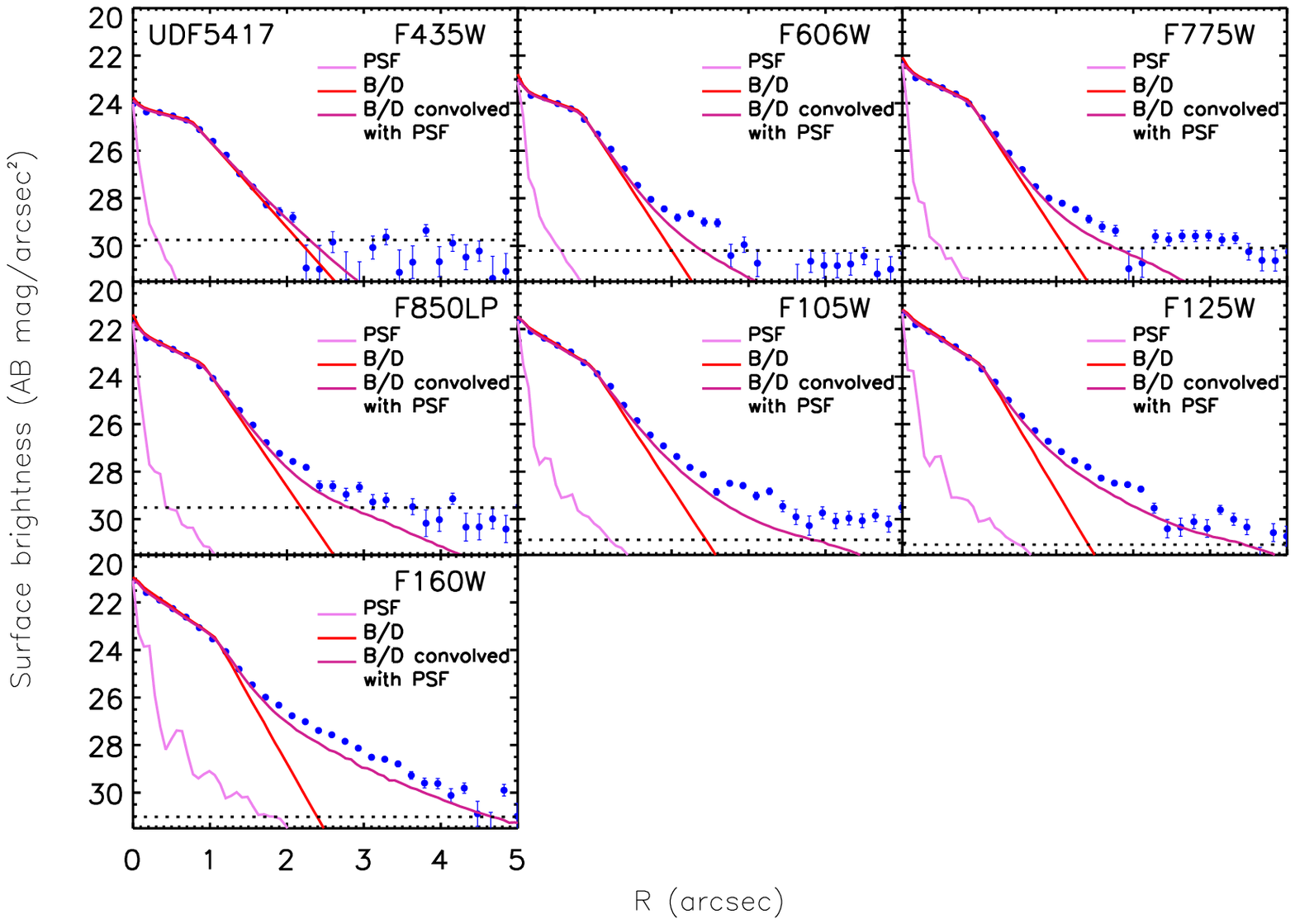}

\caption{Surface brightness profile of the UDF5417 galaxy in all the observed bands. The PSF profiles normalized to the same central surface brightness of the galaxy, as well as the galaxy models plus the galaxy models convolved with the PSFs are shown. The dashed lines in both panels indicate the limiting surface brightness down to which the surface brightness profiles are reliable.} 

\label{psfeffect5417} 
\end{figure*}

We show in this section the effect of the PSF on the surface brightness profiles of both galaxies and
in all the bands used in this work. For the galaxy UDF3372, we show the effect of the PSF in Fig.
\ref{psfeffect3372}, and for the galaxy UDF5417 in Fig. \ref{psfeffect5417}.

It is worth noting that the excess of light above the PSF model prediction is observed in all the
profiles except in the F160W NICMOS profile of the galaxy UDF3372. The effect of the PSF in that band
is so large that does not permit exploring any potential stellar halo light at those distances. Also,
in the F435W band, in both galaxies, the data is not deep enough to explore the presence of a stellar
halo at those (ultraviolet rest-frame) wavelengths.

\section{Bulge-disc-stellar halo decompositions}

\begin{figure*}
\includegraphics[width=\textwidth]{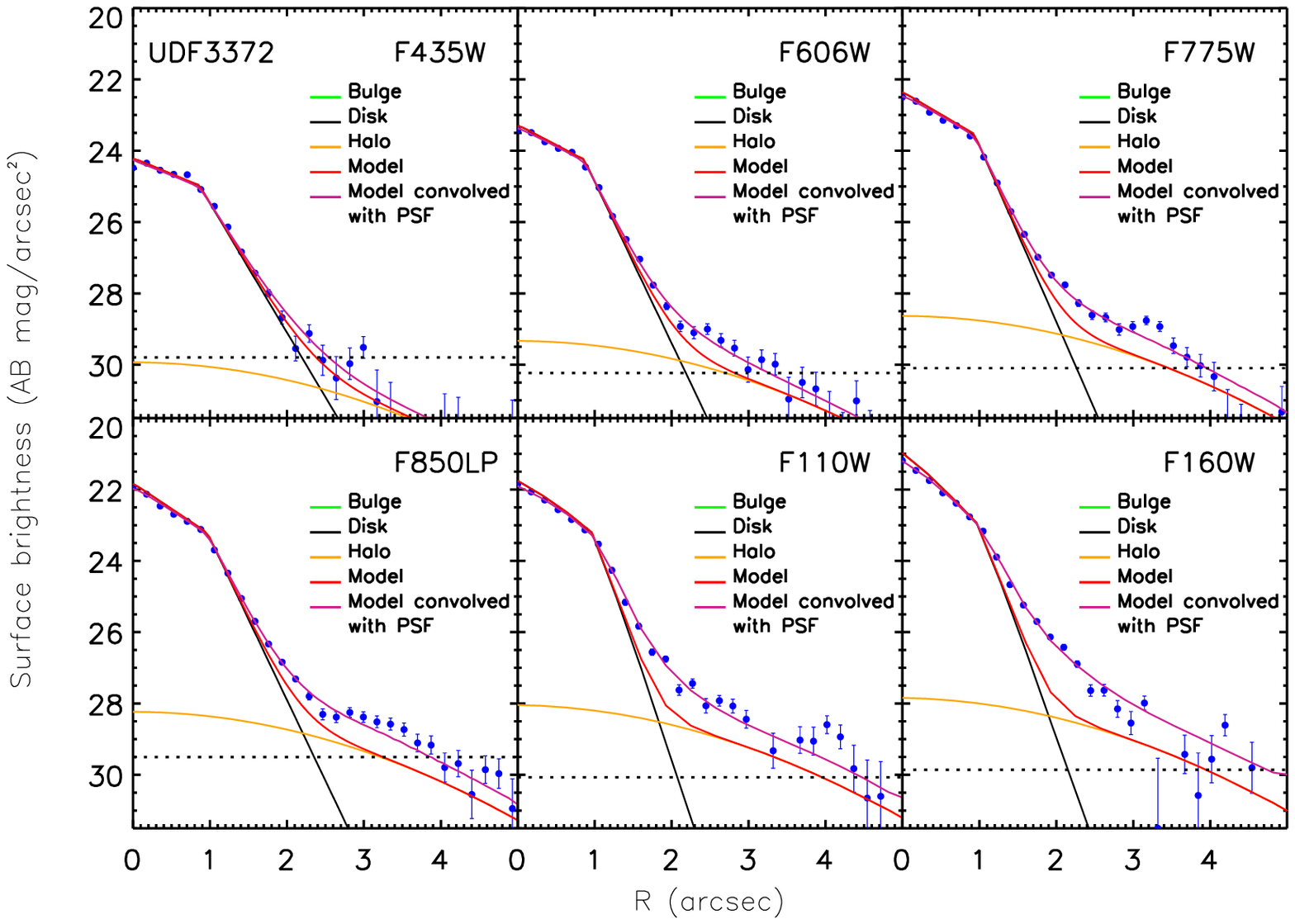}

\caption{Bulge-disc-halo decomposition of the UDF3372 galaxy in all the observed
bands. The bulge, disc, halo as well as the galaxy models plus the galaxy models
convolved with the PSFs are shown. The dashed lines in both panels indicate the
limiting surface brightness down to which the surface brightness profiles are
reliable.} 

\label{bulgedischalo3372} 
\end{figure*}

\begin{figure*}
\includegraphics[width=\textwidth]{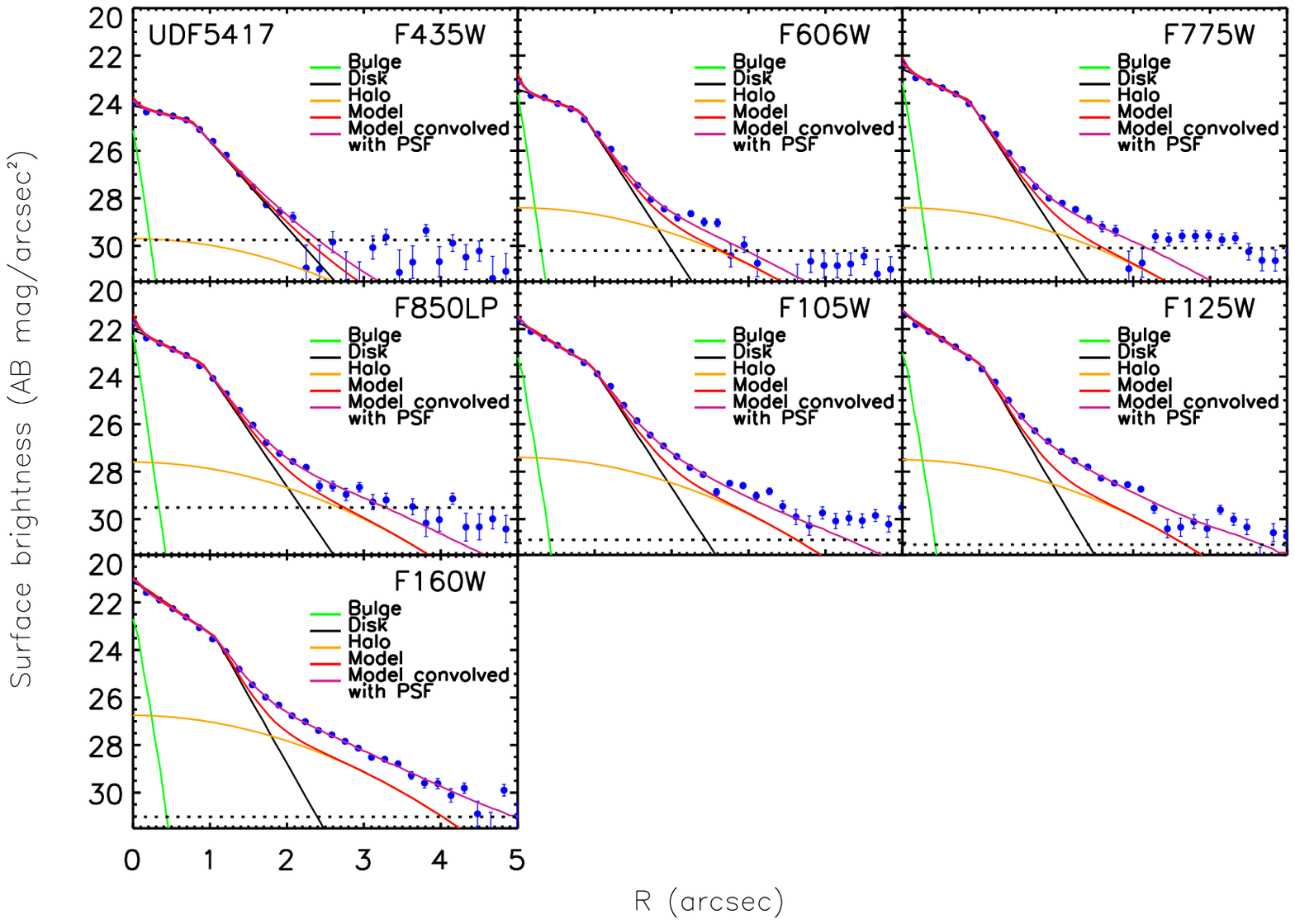}

\caption{Bulge-disc-halo decomposition of the UDF5417 galaxy in all the observed
bands. The bulge, disc, halo as well as the galaxy models plus the galaxy models
convolved with the PSFs are shown. The dashed lines in both panels indicate the
limiting surface brightness down to which the surface brightness profiles are
reliable.} 

\label{bulgedischalo5417} 
\end{figure*}

In this Appendix, we show the bulge-disc-stellar halo decomposition for the two galaxies explored in
this work in all the bands that we have used in this paper. The results of doing this decomposition
are presented in Fig. \ref{bulgedischalo3372} for UDF3372 and Fig. \ref{bulgedischalo5417} for
UDF5417. 

\section{Results based on modeling the halo with an exponential profile}

In this work, we have assumed that the surface brightness distribution of the stellar halo follows a
Gaussian distribution (i.e. a S\'ersic model with a n index fixed to n=0.5). This is done for two
reasons: first, it produces a fit in all the bands which agrees very well with the data distribution, and
secondly, it allow us to compare with the results obtained in the local sample by Bakos \& Trujillo
(2012), where this model was used to extract the properties of the stellar haloes. Nonetheless, the
exact determination of the shape of the stellar halo is uncertain due to the limited spatial range
that we can cover with our observations (not reliable beyond 40 kpc). For that reason, we have
repeated our full analysis based on an exponential distribution for the outer component. Some of the
values obtained directly from this analysis can not be straightforward compared with the local sample
by Bakos \& Trujillo (2012) as those values (for instance: the mean surface brightness, the effective
radius of the halo, etc.) strongly depend on the model used to fit the outer component (see e.g. table
3 of Gilbert et al. 2012). However, other parameters such as the fraction of light in the stellar halo
compared to the total luminosity of the galaxy and/or the colour-corrected profiles help to probe how
dependent are our results on the chosen shape for the stellar outer component.

Exponential profiles for fitting the outer component of the stellar halo produce larger effective
radii and fainter surface brightnesses  (UDF3372: R$_{e,halo}$=47.2$\pm$5.7  kpc,
$<$$\mu_{e,halo}$$>$=26.4$\pm$0.3 mag/arcsec$^2$ (r-band restframe) ; UDF5417:
R$_{e,halo}$=44.1$\pm$2.2  kpc,  $<$$\mu_{e,halo}$$>$=26.4$\pm$0.1  mag/arcsec$^2$ (r-band rest frame))
that in the case of Gaussian models. The stellar halo light contributions using exponential model
raise to 9.1$\pm$2.3 per cent (UDF3372) and 6.5$\pm$0.7 per cent (UDF5417). This is understable due to the larger
amount of light that is located on the tail of an exponential model compared to a Gaussian
distribution. Nonetheless, in the case of UDF5417, where our data are most reliable and deep, the
stellar halo light contribution using both models are relatively similar.

\begin{figure*}
\includegraphics[width=\textwidth]{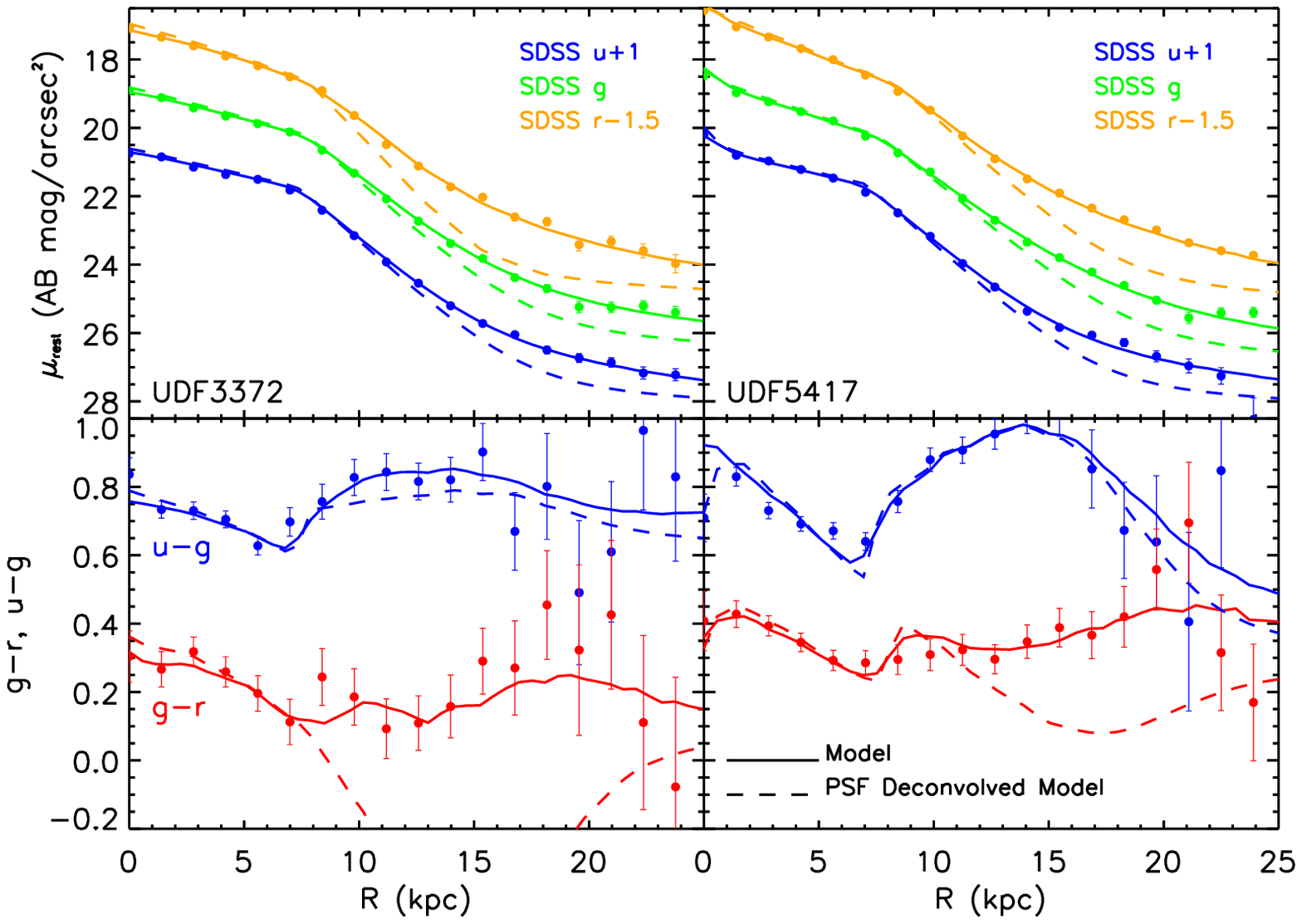}

\caption{{\it Upper panels:} the u, g and r-band equivalent rest-frame surface brightness profiles obtained by the interpolation of the
 observed band profiles.  The solid lines represent the rest-frame surface brightness profiles obtained by the interpolation of the model fits to the observed band profiles. The dashed lines represent the rest-frame surface brightness profiles obtained by the interpolation of the PSF deconvolved model fits to the observed bands. The rest-frame surface brightness profiles have been corrected by the cosmological dimming. {\it Bottom panels:} The g-r and u-g restframe color profiles are shown. The solid and dashed lines are the colours obtained from the model fits to the data (solid line) and the PSF deconvolved models (dashed lines). The stellar haloes have been modelled using exponential profiles.} 

\label{colorexponential} 
\end{figure*}

Finally, we have explored how the colour correction we have applied due to the effect of the PSF in our
surface brightness profiles is altered when we use an exponential model. In Fig.
\ref{colorexponential}, we show the u, g and r rest-frame surface brightness profiles as well as the u-g
and g-r colors PSF corrected radial profiles when using an exponential model to fit the outer region.
That figure shows that, within the error bars, the PSF color corrected radial profiles are not
significantly different assuming a Gaussian or an exponential model. 

\end{document}